\documentclass[12pt]{article}
\usepackage{amsfonts}
\usepackage{amsmath}
\usepackage{latexsym}
\usepackage{bbm}

\def\d{\delta}

\def\m{\mu}

\def\L{\Lambda}

\def\pa{\partial}

\def\Tr{{\rm Tr}}
\def\and{{\rm and}}

\begin{document}
\vspace*{-.6in} \thispagestyle{empty}
\begin{flushright}
CALT-68-2689
\end{flushright}
\baselineskip = 18pt

\vspace{1.5in} {\Large
\begin{center}
Ghost-Free Superconformal Action for Multiple M2-Branes\end{center}}

\begin{center}
Miguel A. Bandres, Arthur E. Lipstein and John H. Schwarz
\\
\emph{California Institute of Technology\\ Pasadena, CA  91125, USA}
\end{center}
\vspace{1in}

\begin{center}
\textbf{Abstract}
\end{center}
\begin{quotation}
\noindent The Bagger--Lambert construction of ${\cal N} =8$
superconformal field theories (SCFT) in three dimensions is based on
3-algebras. Three groups of researchers recently realized that an
arbitrary semisimple Lie algebra can be incorporated by using a
suitable Lorentzian signature 3-algebra. The $SU(N)$ case is a
candidate for the SCFT describing coincident M2-branes. However,
these theories contain ghost degrees of freedom, which is
unsatisfactory. We modify them by gauging certain global symmetries.
This eliminates the ghosts from these theories while preserving all
of their desirable properties. The resulting theories turn out to be
precisely equivalent to ${\cal N}=8$ super Yang--Mills theories.
\end{quotation}

\newpage

\pagenumbering{arabic}

\section{Introduction}

Bagger and Lambert \cite{Bagger:2006sk,Bagger:2007jr,Bagger:2007vi},
as well as Gustavsson \cite{Gustavsson:2007vu,Gustavsson:2008dy}
discovered the general rules for constructing an action for a
three-dimensional theory with $OSp(8|4)$ superconformal symmetry.
Their solution is based on a 3-algebra, which is characterized by
structure constants $f^{ABC}{}_D$ and a metric $h_{AB}$. The initial
assumption was that the metric should be positive definite. This led
to the discovery of a theory with $SO(4)$ gauge symmetry
\cite{Bagger:2007jr}. Its full superconformal symmetry was verified
in \cite{Bandres:2008vf}, which also conjectured its uniqueness. The
uniqueness of this theory was proved in
\cite{Papadopoulos:2008sk,Gauntlett:2008uf}. A proposal for its
physical interpretation in terms of M2-branes in M-theory at an
M-fold singularity has been given in
\cite{Mukhi:2008ux,Distler:2008mk}.

These developments left unresolved the question whether it is
possible to give a Lagrangian description of the conformal field
theory associated with coincident M2-branes in flat 11-dimensional
spacetime. That theory is known to correspond to the IR fixed point
of ${\cal N}=8$ super Yang--Mills theory. The question is whether
there is a dual formulation of this fixed-point theory. The only
apparent way of evading the uniqueness theorem is to consider
3-algebras with an indefinite signature metric. This possibility was
examined by three different groups
\cite{Gomis:2008uv,Benvenuti:2008bt,Ho:2008ei}, who proposed a new
class of theories based on a 3-algebra with Lorentzian signature.
The generators of the 3-algebra are the generators of an arbitrary
semisimple Lie algebra plus two additional null generators $T^\pm$.
The theory based on the 3-algebra associated to the gauge group
$SU(N)$ or $U(N)$ looks like a good candidate for the theory of $N$
coincident M2-branes, except for the fact that it contains unwanted
negative norm states in the physical spectrum. This makes the theory
nonunitary even though these states do not contribute to loops.
Subsequent papers discussing the interpretation and application of
Lorentzian 3-algebras include \cite{Morozov:2008rc} --
\cite{FigueroaO'Farrill:2008zm}. In particular,
\cite{FigueroaO'Farrill:2008zm} proved that the Lorentzian
3-algebras considered in
\cite{Gomis:2008uv,Benvenuti:2008bt,Ho:2008ei} are the {\it only}
indecomposable Lorentzian 3-algebras (aside from the obvious
$SO(3,1)$ variant of the Bagger--Lambert theory).

In this paper we propose modifying the construction in
\cite{Gomis:2008uv,Benvenuti:2008bt,Ho:2008ei} by gauging certain
global symmetries.\footnote{After this work had been completed,
Hirosi Ooguri informed us that Masahito Yamazaki is also considering this
possibility.} We claim that this eliminates the unwanted ghost
degrees of freedom while preserving all of the other symmetries. In
Section 2 we explain the basic idea of our construction in a
simplified model. Section 3 applies the same procedure to the theory
of interest.

\section{The Basic Idea}

After integrating out certain auxiliary fields, the theory proposed
in \cite{Gomis:2008uv,Benvenuti:2008bt,Ho:2008ei} contains terms of
the form
$$
S \sim \int d^3x \left( -\phi_+^{-2} \Tr (F^2)
+ \pa^\m \phi_+  \pa_\m \phi_-\right)
$$
This has manifest scale invariance if $\phi_{\pm}$ have dimension $1/2$.
This theory has a ghost degree of freedom, which (ignoring the first
term) is reminiscent of the one contained in the covariant
gauge-fixed string world-sheet theory prior to imposing the Virasoro
constraints. In the present case, there are no Virasoro constraints,
so the theory needs to be modified if we wish to make sense of it.

An important clue is that this theory has a global symmetry given by
a constant shift of the field $\phi_-$. Our proposal is to
modify this theory by gauging this symmetry through the inclusion of
a dimension $3/2$ St\"uckelberg field $C_\m$
\[
S \sim \int d^3x \left( -\phi_+^{-2} \Tr (F^2) + \pa^\m \phi_+
(\pa_\m \phi_- - C_\m)\right) .
\]
The gauge symmetry is simply given by
$$
\d \phi_- = \Lambda \quad \and \quad \d C_\m = \pa_\m \Lambda.
$$
Classically, this theory is conformally invariant. (In the case of
the M2-brane theory in the next section the conformal symmetry is
expected to survive in the quantum theory.) This theory can be gauge
fixed by setting $\phi_- =0$. Integrating out $C_\m$ gives a
delta functional imposing the constraint $\pa_\m \phi_+ =0$. Thus,
$\phi_+$ is a constant, which is determined by a boundary condition.
Calling the constant $g_{\rm YM}$, we are left with pure Yang--Mills
theory
$$
S \sim -g_{\rm YM}^{-2} \int d^3x  \Tr (F^2) .
$$

The Yang--Mills theory is not conformally invariant, of course,
since $g_{\rm YM}$ is dimensionful. However, this construction shows
that it arises from spontaneous breaking of the conformal symmetry.

\section{Modifying the BL Theory}

Using the notation of \cite{Benvenuti:2008bt}, we start with the
following Bagger--Lambert theory based on a family of 3-algebras
with Lorentzian metric:
\begin{multline}
\mathcal{L} =-\frac{1}{2}\mathrm{Tr}\left( D_{\mu }X^{I}D^{\mu
}X^{I}\right) +D_{\mu }X_{+}^{I}D^{\mu
}X_{-}^{I}+\frac{i}{2}\mathrm{Tr} \left( \bar{\Psi}\Gamma ^{\mu
}D_{\mu }\Psi \right) -\frac{i}{2}\bar{\Psi}_{+}\Gamma ^{\mu }D_{\mu
}\Psi _{-}-\frac{i}{2}\bar{
\Psi}_{-}\Gamma ^{\mu }D_{\mu }\Psi _{+}  \\
+\epsilon ^{\mu \nu \lambda }\mathrm{Tr}\left( \mathcal{B}_{\lambda
}\left( \partial _{\mu }\mathcal{A}_{\nu }-\left[ \mathcal{A}_{\mu
}, \mathcal{A}_{\nu }\right] \right) \right)
-\frac{1}{12}\mathrm{Tr}\left( X_{+}^{I}\left[ X^{J},X^{K}\right]
+X_{+}^{J}\left[ X^{K},X^{I}\right] +X_{+}^{K}\left[
X^{I},X^{J}\right]
\right) ^{2}  \\
+\frac{i}{2}\mathrm{Tr}\left( \bar{\Psi}\Gamma _{IJ}X_{+}^{I}\left[
X^{J},\Psi \right] \right) +\frac{i}{4}\mathrm{Tr}\left(
\bar{\Psi}\Gamma _{IJ}\left[ X^{I},X^{J}\right] \Psi _{+}\right)
-\frac{i}{4}\mathrm{Tr} \left( \bar{\Psi}_{+}\Gamma _{IJ}\left[
X^{I},X^{J}\right] \Psi \right) , \label{Lagrangian}
\end{multline}
where $I=1,...,8$ are the transverse coordinates and
$X_{\pm}^{I}=\frac{1}{\sqrt{2}}\left(X_{0}^{I}\pm X_{1}^{I}\right)$.
The covariant derivatives are defined as
\begin{subequations}
\begin{eqnarray}
D_{\mu }X^{I} &=&\partial _{\mu }X^{I}-2\left[ \mathcal{A}_{\mu
},X^{I}
\right] -\mathcal{B}_{\mu }X_{+}^{I},  \label{Cderivatives} \\
D_{\mu }X_{-}^{I} &=&\partial _{\mu }X_{-}^{I}-\mathrm{Tr}\left(
\mathcal{B}
_{\mu }X^{I}\right) , \\
D_{\mu }X_{+}^{I} &=&\partial _{\mu }X_{+}^{I}
\end{eqnarray}
\end{subequations}
and similarly for the fermions. Note that this theory has a
noncompact gauge group whose Lie algebra is a semidirect sum of any
ordinary Lie algebra $\mathfrak{g}$ of a compact Lie group
$\mathcal{G}$, and dim($\mathfrak{g}$) abelian generators. The gauge
field $\mathcal{A}_{\mu }$ is associated with the compact part,
while the gauge field $\mathcal{B}_{\mu }$ is associated with the
noncompact part. This theory was recently proposed in
\cite{Gomis:2008uv,Benvenuti:2008bt,Ho:2008ei}. Various details of
this Lagrangian, including its field content, gauge symmetry, and
supersymmetry transformations, are given in the Appendix. Like all
BL theories, it has $\mathcal{N}=8$ supersymmetry, scale invariance,
conformal invariance, and $SO(8)$ $R$-symmetry. These combine to
give the supergroup $OSp(8|4)$. The theory also has parity
invariance. At the same time, it does not admit any tunable coupling
constant, since any coupling constant can be absorbed in field
redefinitions. Furthermore $\mathcal{G}$ can be chosen to be any
compact Lie group. These are special features that are not shared by
the $SO(4)$ BL theory, which is based on a 3-algebra with a
positive-definite metric.

Despite the numerous properties which make this theory a promising
candidate for describing multiple M2-branes in flat space, it has
one very troubling feature. To see this, consider the fields
$X_{-}^{I}$ and $\Psi _{-}$. Note that the full dependence on these
fields is given by:
\begin{equation}
\mathcal{L}_{-}=-i\bar{\Psi}_{+}\Gamma ^{\mu }\partial _{\mu }\Psi
_{-}+\partial ^{\mu }X_{+}^{I}\partial _{\mu }X_{-}^{I}.
\label{LGhost}
\end{equation}
As it stands, these terms describe propagating ghost degrees of
freedom, which makes the theory unsatisfactory, since it is not
unitary.  At this point, it is useful to observe that the action has
the following global shift symmetries (pointed out in
\cite{Benvenuti:2008bt}):
\begin{equation*}
\delta X_{-}^{I}=\Lambda ^{I}\quad \mathrm{and}\quad \delta \Psi
_{-}=\eta .
\end{equation*}
Also note that $\Psi _{-}$ and $X_{-}^{I}$ do not appear in any of
the gauge or SUSY transformations of the other fields. We will show
that it is possible to eliminate the ghosts from the theory, while
preserving all of its desirable properties, by promoting these
global shift symmetries to local symmetries.

To gauge the global shift symmetries described above we introduce
two new gauge fields: a vector field $C_{\mu}^{I}$ in the vector
representation of $ SO(8)$, and a 32-component Majorana--Weyl spinor
$\chi$ satisfying $\Gamma ^{012}\chi =-\chi $. These appear in two
new terms which we add to the Lagrangian:
\begin{equation}
\mathcal{L}_{\mathrm{new}}=\bar{\Psi}_{+}\chi -\partial ^{\mu
}X_{+}^{I}C_{\mu }^{I}.  \label{Lnew}
\end{equation}
Note that $C_{\mu}^{I}$ must have dimension 3/2 and $\chi $ must
have dimension 2 to preserve scale invariance. The new local shift
symmetries are
\begin{equation}
\delta X_{-}^{I}=\Lambda ^{I},\quad \delta C_{\mu }^{I}=\partial
_{\mu }\Lambda ^{I}  \label{gtX}
\end{equation}
and
\begin{equation}
\delta \Psi _{-}=\eta ,\quad \delta \chi =i\Gamma ^{\mu }\partial
_{\mu }\eta .  \label{gtP}
\end{equation}
There is one additional local symmetry of Eq.~(\ref{Lnew}), which is
relatively trivial, namely
\begin{equation}
\delta C_{\mu }^{I}=\partial ^{\rho }\tilde{\Lambda}_{\mu \rho
}^{I},\quad \mathrm{where}\quad \tilde{\Lambda}_{\mu \rho
}^{I}=-\tilde{\Lambda}_{\rho \mu }^{I}.  \label{trivial}
\end{equation}
$C_{\mu }^{I}$ and $\chi $ are invariant under the original gauge
symmetries.

Now let us consider the supersymmetry of the modified theory. The
supersymmetry transformations of all the old fields are unchanged.
In particular,
\begin{equation}
\delta X_{+}^{I}=i\bar{\varepsilon}\Gamma ^{I}\Psi _{+}
\end{equation}
and
\begin{equation}
\delta \Psi _{+}=\Gamma ^{\mu }\partial _{\mu }X_{+}^{I}\Gamma
^{I}\varepsilon .
\end{equation}
The supersymmetries of the new gauge fields must be defined in such
a way that $\mathcal{L}_{\mathrm{new}}$ is invariant. We will find
that the resulting supersymmetry algebra closes on shell when one
takes account of the new gauge symmetries. Under supersymmetry
\begin{equation}
\delta C_{\mu }^{I}=\bar{\varepsilon}\Gamma ^{I}\Gamma _{\mu }\chi
\end{equation}
and
\begin{equation}
\delta \chi =i\Gamma ^{I}\varepsilon \,\partial ^{\mu }C_{\mu }^{I}.
\end{equation}
Using these four transformation rules, it is easy to see that both
$\mathcal{L}_{ \mathrm{new}}$ and the equations of motion are
supersymmetric.

We will now check the closure of all the algebras. The fact that the
supersymmetry variations of $C_{\mu}^{I}$ and $\chi $ are not
invariant under the new gauge transformations implies
that the supersymmetry transformations do not
commute with these gauge transformations. Specifically, one finds that
\begin{equation}
\lbrack \delta (\Lambda ),\delta (\varepsilon )]=\delta (\eta
),\quad \mathrm{where}\quad \eta =\Gamma ^{\mu }\Gamma ^{I}\partial
_{\mu }\Lambda ^{I}\varepsilon
\end{equation}
and
\begin{equation}
\lbrack \delta (\eta ),\delta (\varepsilon )]=\delta (\Lambda
)+\delta ( \tilde{\Lambda})\quad \mathrm{where}\quad \Lambda
^{I}=i\bar{\varepsilon} \Gamma ^{I}\eta \quad \mathrm{and}\quad
\tilde{\Lambda}_{\mu \rho }^{I}=i \bar{\varepsilon}\Gamma ^{I}\Gamma
_{\mu \rho }\eta .
\end{equation}
The supersymmetry algebra is slightly affected, as well.
Specifically, we find that
\begin{equation}
\lbrack \delta (\varepsilon _{1}),\delta (\varepsilon _{2})]C_{\mu
}^{I}=\delta (\mathbf{\xi })C_{\mu }^{I}+\delta
(\tilde{\Lambda})C_{\mu }^{I},
\end{equation}
where $\xi ^{\rho }=2i\bar{\varepsilon}_{1}\Gamma ^{\rho
}\varepsilon _{2}$, as usual, and $\tilde{\Lambda}_{\mu \rho
}^{I}=\xi _{\mu }C_{\rho }^{I}-\xi _{\rho }C_{\mu }^{I}$. Similarly,
for $\chi $ we find that
\begin{equation}
\lbrack \delta (\varepsilon _{1}),\delta (\varepsilon _{2})]\chi
=\delta ( \mathbf{\xi })\chi +\delta (\eta )\chi ,
\end{equation}
where
$\eta=\left(-\bar{\epsilon_{1}}\Gamma^{\mu}\epsilon_{2}\Gamma_{\mu}
+\frac{1}{4}\bar{\epsilon}_{1}\Gamma^{LM}\epsilon_{2}\Gamma_{LM}\right)\chi$.
One also finds that requiring the on-shell closure of the commutator
$[\delta (\varepsilon _{1}),\delta (\varepsilon _{2})]\Psi _{-}$
gives the expected equation of motion for $\Psi _{-}$ after noting
that the commutator receives a contribution from $\delta (\eta )\Psi
_{-}$. In summary, we have verified that the supersymmetries close
on shell into translations, the old gauge transformations, and the
new gauge transformations given by Eqs~(\ref{gtX})--(\ref{trivial}).

\section{Discussion}

After modifying the theory by introducing the new gauge fields
$C_{\mu}$ and $\chi$, it still has scale invariance, $\mathcal{N}=8$
supersymmetry, no coupling constant, and can accommodate any Lie
group in its gauge group, which are all desirable properties for
describing multiple M2-branes in flat space. In addition, we can use
the new gauge symmetries to make the gauge choices
\[
X_{-}^{I}=\Psi_{-}=0.
\]
This removes the kinetic terms for the ghosts and changes the
supersymmetry transformations for $C_{\mu}$ and $\chi$ by induced
gauge transformations, i.e. $\delta
C_{\mu}^{I}=\bar{\epsilon}\Gamma^{I}\Gamma_{\mu}\chi+\partial_{\mu}\Lambda^{I}$
and $\delta\chi=i\Gamma^{I}\epsilon\partial^{\mu}C_{\mu}^{I}
+i\Gamma^{\mu}\partial_{\mu}\eta$ for appropriate choices of $\L^I$
and $\eta$. Furthermore, the equations of motion that come from
varying the new fields are
\[
\partial_{\mu}X_{+}^{I}=0,\,\,\,\Psi_{+}=0.
\]
The first equation implies that the $X_{+}^{I}$ is a constant. Any
nonzero choice spontaneously breaks conformal symmetry and breaks
the R-symmetry to an unbroken $SO(7)$ subgroup.
On the other hand, the choice $X_{+}^{I}=0$ gives a free theory.

We can use the $SO(8)$ R-symmetry to choose the nonzero component of
$X_+^I$ to be in the 8 direction, $X_{+}^{I}=v\delta^{I8}$. Also,
the noncompact gauge fields, ${\mathcal B}$, which appear
quadratically can be integrated out. This leaves a maximally
supersymmetric 3d Yang-Mills theory with $SO(7)$ R-symmetry:
\[
\mathcal{L}=-\frac{1}{4v^{2}}{\normalcolor
\Tr}\left(F_{\mu\nu}F^{\mu\nu}\right)-\frac{1}{2}{\normalcolor
\Tr}\left(D'_{\mu}X^{i}D'_{\mu}X^{i}\right)+\frac{i}{2}{\normalcolor
\Tr\left(\bar{\Psi}\Gamma^{\mu}D'_{\mu}\Psi\right)}\]
\[
+\frac{i}{2}{\normalcolor
\Tr}\left(\bar{\Psi}\Gamma_{8i}\left[X^{i},\Psi\right]\right)
-\frac{v^{2}}{4}{\normalcolor
\Tr}\left(\left[X^{i},X^{j}\right]\right)^{2}\] where the index
$i=1,...,7$, and $D'_{\mu}$ and $F_{\mu\nu}$ depend only the
massless gauge field $\mathcal{A}$ associated with the maximally
compact subgroup of the original gauge group. Note that this is an
exact result -- not just the leading term in a large-$v$ expansion.
This is a supersymmetric generalization of the toy model described
in Section 2.

To summarize, in this paper we have proposed a modification of the
Bagger-Lambert theory that removes the ghosts when the 3-algebra has
a Lorentzian signature metric, thus ensuring unitarity. Such
theories evade the no-go theorem, which states that there is
essentially only one nontrivial 3-algebra with positive-definite
metric. Our modification of the Lorentzian 3-algebra theories in
\cite{Gomis:2008uv,Benvenuti:2008bt,Ho:2008ei} breaks the conformal
symmetry spontaneously and reduces them to maximally supersymmetric
3d Yang-Mills theories.\footnote{Reference \cite{Ho:2008ei} observed
that if one chooses $X_+^I$ to be constant and $\Psi_+$ to be zero,
then the theory reduces to N=8 SYM. However, they did not deduce
these choices from an action principle.} This result is somewhat
disappointing inasmuch as it means that we are no closer to the
original goal of understanding the $v\to \infty$ IR fixed-point
theory that describes coincident M2-branes in 11 noncompact
dimensions. As things stand, it appears that the BL $SO(4)$ theory
is the only genuinely new maximally supersymmetric superconformal
theory. Of course, one should still explore whether there are other
3-algebras (whose metric is neither positive-definite not
Lorentzian) that open new possibilities.

Note added: After this paper was first posted, two related papers
appeared \cite{Gomis:2008be,Ezhuthachan:2008ch}. Also, a paper by
Aharony et al. appeared that introduces a very promising class of
theories with ${\cal N} =6$ superconformal symmetry
\cite{Aharony:2008ug}. It proposes that these theories actually have
${\cal N} =8$ superconformal symmetry (implemented in a very subtle
manner) in the appropriate cases.

\section*{Acknowledgments}

We have benefitted from discussions with Joe Marsano. This work was
supported in part by the U.S. Dept. of Energy under Grant No.
DE-FG03-92-ER40701.


\section*{Appendix. BL Theory for General Lie Algebras}

In this appendix, we follow the notation of \cite{Benvenuti:2008bt}.
The Lagrangian of a \textrm{BL}-theory is completely specified once
a 3-algebra with a metric is given. The structure constants of the
3-algebra $ f^{ABC}{}_{D}$ must satisfy the fundamental identity and
$f^{ABCE}=f^{ABC}{}_{D}h^{DE} $, where $h^{DE}$ is the 3-algebra
metric, must be totally antisymmetric. In \cite{Benvenuti:2008bt},
the 3-algebra is constructed from an ordinary Lie algebra
$\mathfrak{g}$ by adding two generators to $\mathfrak{g}$ called $T^+$
and $T^-$ so that the 3-algebra has dimension $dim\left(
\mathfrak{g}\right) +2$. Its structure constants are given in terms
of the $\mathfrak{g}$-structure constants $f^{ab}{}_{c}$ as
\begin{equation}
f^{+ab}{}_{c}=f^{ab}\!_{c}, \label{FI}
\end{equation}
with all other nonzero components of $f^{ABC}{}_{D}$ related by
permuting, raising, or lowering indices. The generators of
$\mathfrak{g}$ satisfy
\begin{eqnarray}
\left[ T^{a},T^{b}\right]  &=&f^{ab}\!_{c}T^{c},  \label{algebra} \\
\mathrm{Tr}\left( T^{a}T^{b}\right)  &=&\delta ^{ab}.  \notag
\end{eqnarray}
The invariant metric of the 3-algebra is given by
\begin{equation}
h^{+-}=-1,\qquad h^{++}=0,\qquad h^{--}=0,\qquad h^{ab}=\delta
^{ab}. \label{metric}
\end{equation}

With the choice of structure constants and 3-algebra metric given
above, the BL theory reduces to the Lagrangian given in Eq.~1. The
field content of the theory is summarized in the following table.

\begin{table}[hbt]
\centering
\begin{tabular}{ccccc}
\hline
\text{Field} & 3d\text{ World Volume} & SO(8) & $\mathfrak{g}$ & \text{Dimension} \\
\hline
$X_{\pm }^{I}$ & \text{Scalar} & $8_{\mathrm{v}}$ & Singlet & 1/2 \\
$X^{I}$ & \text{Scalar} & $8_{\mathrm{v}}$ & \text{Adjoint} & $1/2$ \\
$\Psi _{\pm }$ & $\text{Spinor}$ & $8_{\mathrm{s}}$ & Singlet & 1 \\
$\Psi$  & $\text{Spinor}$ & $8_{\mathrm{s}}$ &
\text{Adjoint} & 1
\\
$\mathcal{A}_{\mu }$ & \text{Gauge field} & 1 & \text{Adjoint} & 1 \\
$\mathcal{B}_{\mu }$ & \text{Gauge field} & 1 & \text{Adjoint} & 1 \\
\hline
\end{tabular}
\end{table}

The gauge transformations are
\begin{subequations}
\begin{eqnarray}
\delta X^{I} &=&2\left[ \Lambda ,X^{I}\right] +MX_{+}^{I},  \label{GT} \\
\delta X_{-}^{I} &=&\mathrm{Tr}\left( MX^{I}\right) , \\
\delta X_{+}^{I} &=&0, \\
\delta \Psi  &=&2\left[ \Lambda ,\Psi \right] +M\Psi _{+}, \\
\delta \Psi _{-} &=&\mathrm{Tr}\left( M\Psi\right) , \\
\delta \Psi _{+} &=&0. \\
\delta \mathcal{A}_{\mu } &=&\partial _{\mu }\Lambda +2\left[
\Lambda ,
\mathcal{A}_{\mu }\right] , \\
\delta \mathcal{B}_{\mu } &=&\partial _{\mu }M+2\left[
M,\mathcal{A}_{\mu } \right] +2\left[ \Lambda ,\mathcal{B}_{\mu
}\right] ,
\end{eqnarray}
where $\Lambda $ and $M$ are infinitesimal matrices in the adjoint of
$\mathfrak{g}$. The matrix $\Lambda $ generates the $\mathcal{G}$
gauge transformations while $M$ generates the noncompact subgroup
transformations.

Finally, the $\mathcal{N}=8$ SUSY transformations (consistent with
scale invariance) are
\end{subequations}
\begin{subequations}
\begin{eqnarray}
\delta \mathcal{A}_{\mu } &=&\frac{i}{2}\bar{\varepsilon}\Gamma _{\mu }\Gamma
_{I}\left( X_{+}^{I}\Psi -X^{I}\Psi _{+}\right) ,  \label{susy} \\
\delta \mathcal{B}_{\mu } &=&i\bar{\varepsilon}\Gamma _{\mu }\Gamma
_{I}\left[ X^{I},\Psi \right] ,\\
\delta X_{\pm }^{I} &=&i\bar{\varepsilon}\Gamma ^{I}\Psi _{\pm }, \\
\delta X^{I} &=&i\bar{\varepsilon}\Gamma ^{I}\Psi , \\
\delta \Psi _{+} &=&\partial _{\mu }X_{+}^{I}\Gamma ^{\mu }\Gamma
^{I}\varepsilon , \\
\delta \Psi _{-} &=&D_{\mu }X_{-}^{I}\Gamma ^{\mu }\Gamma
^{I}\varepsilon - \frac{1}{3}\mathrm{Tr}\left(
X^{I}X^{J}X^{K}\right) \Gamma _{IJK}\epsilon ,
\\
\delta \Psi  &=&D_{\mu }X^{I}\Gamma ^{\mu }\Gamma ^{I}\varepsilon
-\frac{1}{2 }X_{+}^{I}\left[ X^{J},X^{K}\right] \Gamma
_{IJK}\epsilon .
\end{eqnarray}
\end{subequations}

\end{document}